\documentclass[12pt]{article}
\usepackage{indentfirst}
\usepackage{a4wide}
\usepackage{longtable}
\title{A new hypothesis of sunspot formation}
\author{V.I. Zhukov\\
Novosibirsk State Technical University, Novosibirsk,\\
 K. Marks Ave., 20, 630092, Russia, vizh@ngs.ru}
\date{}

\begin{document}

\maketitle

\begin{abstract}
The process of sunspot formation is considered with the account of heat effects.
According to the Le Chatelier principle, a local overheating must precede to the cooling
of solar surface in the places of sunspot formation. The sunspot dynamics is a process
close to the surface nucleate-free boiling in a thin layer with formation of bubbles (or
craters), so we focus on the analogy between these two processes. Solar spots and surface
nucleate-free boiling in a thin layer have similarities in formation conditions, results
of impact on the surface were they have been formed, periodicity, and their place in the
hierarchy of self-organization in complex systems. The difference is in the working
medium and method of channelling of extra energy from the overheated surface --- for
boiling process, the energy is forwarded to generation of vapor, and in sunspots the
solar energy is consumed to formation of a strong magnetic field. This analogy explains
the problem of a steady brightness (temperature) of a spot that is independent of the
spot size and other
characteristics.\\

\noindent PACS number(s): 96.60.Qc, 68.15.+e, 44.30.+v, 07.30.Cy
\end{abstract}

\section{INTRODUCTION}

Sunspots are observed as moderate-dark formations on the Sun
surface. Today there exist many hypotheses, which are framed from
different premises and able to explain different groups of
features in sunspot generation and its development. Those
hypotheses can be classified into several groups: hydrodynamic,
magnetic, and magnetic-hydrodynamic explanations. Until 30s of the
last century, the hydrodynamic and thermodynamic hypotheses had
been developed; they explained generation of sunspots and their
reduced temperature by adiabatic processes of gas expansion and
cooling. A comprehensive review and analysis of the hydrodynamic
and adiabatic hypotheses were performed by  Sitnik \cite{Sit}.

In 1908, Hall discovered the magnetic field in sunspots. From this
point, magnetic and magnetic-hydrodynamic theories have been
developed. Usually they postulate an existence of general magnetic
field and its amplification in the spot zones due to
self-excitation processes.  The general magnetic field induces the
electric current in a moving ionized gas and this induced field
enhances or attenuates the primary field ("solar hydromagnetic
dynamo" of "dynamo-effect").  Now the existence of a strong
magnetic field is considered as a key feature of sunspots, as a
primary phenomenon that controls their generation, physical and
dynamic characteristics. The review and analysis of magnetic and
magnetic-hydrodynamic theories is available in literature; in
particular, the references to these reviews can be found in
several books (Bray and Loughhead \cite{Bray}; Obridko \cite{Obr})
and in the later papers of other authors.

However, those theories fail to explain some observable phenomena.
According to Obridko \cite{Obr}, there exist some difficulties
concerning energy transfer in a spot and problem of sport
brightness (temperature). The generated spots with penumbras have
a universal brightness (temperature), which is independent of
their size and other parameters. The observed brightness for spots
lies in a very narrow range. The objects with a transitional
brightness are unstable and decay promptly. The balance of
mechanisms providing, on one hand, a reduced spot brightness, but,
on another hand, a non-zero brightness, must be very strict. The
fundamentals of this balance are not clear yet from the
theoretical point of view. All the known mechanisms of development
of the spot magnetic field (diffusion model), spot cooling, and
energy transfer may provide this balance near any state; this
means that we must observe a continuum of spot brightness (or its
temperature).

The magnetic field plays a passive role in photosphere strata and
does not participate directly in energy transfer. It makes a
contribution in a change of substance distribution, and this
changes the conditions for radiation energy transfer. However,
significant concentrations of substance are possible that are
close in temperature, because the magnetic field does not
influence directly the radiation transfer. A spot is a secondary
stable state of solar substance (Obridko \cite{Obr}).

The phenomenon of discretization is very important in energy
aspects. All stable states have different energies, and fast
transitions between them are possible only if matched with
release/consumption of energy that must be significant for
different nonstationary phenomena in solar atmosphere.

There exists a problem of flux deficit in the spot. The problem
can be formulated as follows: the spot umbra produces 15-20\% of
the flux generated by the undisturbed zones of the solar
photosphere. Where does the rest of energy flux difference go:
does it transform into other forms of energy or it is just
distributed uniformly of the solar surface?

Gurevich and Lebedinsky \cite{Gur,Leb} were the first who put forward the idea that the
radiant energy in a spot transforms into the magnetic field energy, but they did not
propose any specific mechanism for that. Later, many researchers tried to explain the
energy flux imbalance by energy transformation into other nontraditional forms: magnetic
field, Evershed effect, motion of magnetic knots, different types of waves, etc.

Besides, we see a problem of effective conversion of heat energy
into ordered mechanical energy. There exists a series of facts
that indicate an intimate connection of sunspots with convection
of different scales: a tendency that sunspots appear at the
junctions of supergranules; a similarity in general view and
properties of light elements in the spot umbra and granulation.

All the magnetic and magnetic-hydrodynamic theories use (in this
or that way) the Lenz's rule: the induced current takes the
direction that its magnetic field counteracts the changes in the
magnetic flux that caused this induction. Actually, the Lenz's
rule is an another presentation of    the Le Chatelier principle
(any system in equilibrium under external impact proceeds an
automatic adjustment to compensate this impact) being applied to a
particular case of electromagnetic phenomena.

The review of literature and problems concerning mechanism of
sunspot formation convinced us that formation of sunspots is
accompanied by heat processes that cannot be accounted by Lenz's
rule; therefore the existing magnetic and magnetic-hydrodynamic
theories fail to explain the effect of constant sunspot
temperature. Obviously, this consideration is incomplete and does
not cover all the sides of this process. If we consider the
evolution of this process with the heat aspects and Le Chatelier
principle, we have to assume that a certain overheating of Sun
surface stimulates the processes that apt to decrease the body
temperature in the given place, and there is a need to channel out
the excessive energy; it is consumed to creation of the magnetic
field in the spot.

The process of liquid boiling when overheat stimulates its cooling
is a close analogy to this process. Cooling takes place due to
formation of a vapor bubble, and the energy goes to intensive
generation of vapor phase.

In this paper we consider an analogy between formation of sunspots
and liquid boiling. We want to apply the expedient that has been
well-proved in physics: the knowledge accumulated in one field is
being transferred to another scope of study and we check up how
common are the regularities in these fields. The considered analogy
must be not contrary to the results of observation.

\section{OBSERVATION OF SUNSPOTS}
The process of sunspot formation proceeds several stages, Obridko
\cite{Obr}: at first, one or several pores develop in a form of
local darkening of intergranulation substance at the junctions of
supergranules with the plumes over these pores. Then the elongated
darkening zones of inter-granular substance develop and the pores
merge. Later a patch of dark substance (or group of pores) starts
condensing and the spot umbra appears. For a large spot, there can
be several of those nucleate umbras. The plumes can be observed
over the spots, but plumes disappear earlier than the spots
dissipate.

The process of sunspot formation starts at a certain temperature
and finishes when a certain level of temperature is reached, i.e.
this is a threshold phenomenon. A fully developed spot has the
sizes multiple to the size of supergranules, and the spots with
intermediate size are usually less stable. Through the
hierarchical system of self-organization of structures, it changes
the regime of natural convection. The magnetic field appears
before the sunspots and disappears after them.

\section{OBSERVATION ON NUCLEATE-FREE BOILING IN A THIN LAYER OF LIQUID}

With a growth in the heat flux, the natural convection mode is
replaced by boiling mode for regular liquids. It is exactly the
boiling mode that can be described by a boiling temperature (at a
given pressure). Studying the mechanism of bubble formation, Moore
and Mesler \cite{Moore} discovered temperature pulsation under the
vapor bubbles. This pulsation develops because for a boiling mode
the local heat flux might be several orders higher than the
general heat flux conveyed by the heating surface; and this causes
a local cooling of the surface.

In the outer manifestation, the process of sunspot formation is quite different from
conventional process of nucleate boiling. According to the Kutateladze's \cite{Kut}
definition, the boiling is a process of evaporation in the liquid bulk. We do not see any
bubbles during sunspot formation; however, the author of this paper had observed the
process of nucleate-free boiling in a thin layer under vacuum - there was no bubbles, but
the funnel-like and crater-like structures were formed which could move over the heating
surface (Gogonin et al., \cite{Gog}; Zhukov \cite{Zh96}; Dorokhov and Zhukov
\cite{Zh99}). The funnel-shape structures might be formed not only in a thin layer over
the heated surface, but also on a surface of overheated thick liquid layer.

The most detail study was performed with a thin oil layer with the depth of 2 mm. Two
cases were studied: with the pressure above the layer equal to 200 Pa and 5-10 Pa. As the
heat flux increases, two processes can be observed: they are different in manifestation,
but the same in their essence.  For the pressure of 200 Pa, a flash boiling took place,
but it was not observed for the pressure of 5-10 Pa, but another process was observed,
with the diagram depicted in Fig. 1.

Fig. 1a shows a cross-section of the convection cell and streamlines. Over the cell, a
sketch of a velocity profile of vapor is depicted. The hot liquid ascends in the center
of cell, and the colder liquid descends at the boundaries. On the free interface of
convection cells, a temperature gradient is usually observed (Berdnikov and Kirdyashkin
\cite{Ber}). As the liquid is heated to the temperature close to the boiling temperature,
it starts to evaporate intensively on the top interface. In this moment, the reactive
force of phase transition has different magnitudes in sites with different heating. In
the sites with hotter liquid it is higher, Kutateladze \cite{Kut}, and its action makes
the layer thinner and produces a ``funnel'' (Fig. 1,b). Existence of ``funnels'' is
supported by the reactive force of the phase transition.

The fact that liquid evaporation rate from the ``funnel'' surface is higher than in other
places was evidenced by visual observation of mist jets rising from a ``funnel'' (it is
visible in rays of light). The hotter zones have more dense population of ``funnels''.
This is also a place for formation of ``craters'' covered with ultra thin layer of oil
(Fig. 1,c).

The observed processes are depicted in Fig. 2. In the place where the crater has passed,
the liquid becomes colder due to intensive evaporation; first one cannot see any ordered
motion of oil flow in this trace, but later convection cells develop up there. After a
time, the ``funnels'' develop on the place of these cells. In the zones with a dense
population of ``funnels'', a crater may be born again. The process tend to reproduce; and
the number, size, and nucleation frequency of ``craters'' become higher with a growth of
the specific heat flux.

But for entire range of heat fluxes, for the regime of joint existence of ``funnels'' and
``craters'', the ``craters'' were always covered with an ultra fine level of oil. At
last, at high specific heat fluxes, one can observe a regime with most of surface covered
with ``craters'' and with moving narrow ``bridges'' of oil moving between them and
providing wetting of ``craters'' (Fig. 2,c). At the heat flux corresponding to the
beginning of this process, the ``craters'' are covered with an ultra thin layer of oil.

In our experiments, we measured the temperature of the heating surface with a
thermocouple placed at the distance of 0.1 mm from the surface. The thermocouple readings
were recorded with a plotter. The evolution of surface temperature under the 2-mm
thickness oil is plotted in Fig. 3 for the specific heat flux $q\sim 10,000$~ W/m$^2$ and
the pressure before heating fixed at the level of 5-10 Pa. At the initial period, there
is no ``craters'' in the oil layer and the temperature of heating surface increases by
exponent (section 1-2).  After $t\sim 15$~sec, ``craters'' develop in the layer and the
surface temperature decreases (section 2-3). Since $t\sim 20$~sec, the temperature of
surface remains almost constant. Since $t\sim 27$~ sec, the thermocouple imbedded into
the bottom demonstrates a passing of a ``crater'', and temperature drops drastically
(section 4-5), and then recovers (section 5-6). The next peak on this plotting
corresponds to passing of another ``crater'' (section 6-7). After $t\sim 40$~ sec, the
temperature of surface again increases exponentially up to the initial level that was in
the system before development of ``craters'' in the layer (section 7-8). The moment of
passing of a ``crater'' over the thermocouple was tracked down by visual observations.

Since the form of event is too different from the regular boiling
of liquids (e.g. no vapor bubbles), the researchers (Gogonin at
al., \cite{Gog}) classified this as an evaporation process, which
takes place indeed. However, we can observe some features inherit
to regular process of nucleate boiling:
\begin{enumerate}
\item  With the growth of heat flux, this regime follows the
regime of convective heat transfer and it is more
intensive.

\item One can observe temperature pulsation of sites under the structural features of
this process (``craters'' and ``funnels'').

\item The surface under the structural features (the basement of ``craters'' and
``funnels'') is covered with an ultra thin layer of liquid.
\end{enumerate}

Taking into consideration these features, the author of the present paper classifies the
process described by Gogonin et al. \cite{Gog} as surface nucleate-free boiling in a thin
layer of liquid in the form of film evaporation. The essence of this mode that a liquid
is being heated through a horizontal surface under vacuum.  Local thinning of the layer
takes place and the structures develop in the form of ``funnels'' (Fig.1,b) and
travelling ``craters'' (Fig. 1,c), which are caused by the impact of reactive force of
phase transition in a non-uniform interface of liquid (Zhukov \cite{Zh96}; Dorokhov and
Zhukov \cite{Zh99}).

\section{COMPARATIVE ANALYSIS OF PHENOMENA}

The table summarizes the comparison between formation of sunspots and process of
nucleate-free boiling in a thin layer of liquid: the dynamics, conditions of formation,
periodicity of processes, result of impact, place in the hierarchy of self-organization
of complex systems, etc.

\begin{longtable}{|p{0.2in}|p{1.75in}|p{1.75in}|p{1.85in}|}
\caption {}\\
 \hline No. & Feature or parameter & Sunspot &
Nucleate-free
boiling\\
\hline 1& 2&3&4\\
\endfirsthead
\caption []{(continued)}\\
\hline 1&2&3&4\\
\endhead
\hline 1 & Dynamics of formation & Generation of one or several
pores fringed with bright strips of flame substance. & Generation
of funnels with mist rising from the surface.\\
\cline{3-4}   & &Merger of pores, formation of spot umbra,
agglomeration into one big spot.&The place with highest population
of funnels generates the crater.\\
\cline{3-4}  & &Ascending of solar substance at the periphery of
solar spots (Evershed's effect).&The jet of vapor rises over the
crater
perimeter from the meniscus zone.\\
\cline{3-4}  & &The spot drifts over the Sun surface.&The crater
travels over the heating surface.\\
\hline  2&Conditions of formation &Under vacuum&Under vacuum\\
\hline 3&Geometry characteristics&Pores and spots are depressions
in the solar photosphere (Willson's effect).&Funnels and craters
are depressions in a thin liquid layer.\\
\hline 4&Sites of generation on the surface&Produced in sites of
local overheating of solar photosphere at the junctions of
supergranules.&Produced in sites of local surface overheating
under the layer of mineral vacuum oil in a terrestrial laboratory
setup.\\
\hline 5&Action on the surface in the formation site&Cooling of
the surface in the formation place. &Cooling of surface under the
formation place.\\
\hline 6&Periodicity&Periodicity feature. The period of formation
is about 11 years.&Periodicity feature The period of formation is
about several seconds.\\
\hline 7&Place in the hierarchy of self-organization of complex
systems&Follows after the regime of natural convection.&Follows
after the regime of natural convection.\\
\hline 8&Influence of pressure&The growing pressure of magnetic
field suppresses development of spot.&The growing pressure of
vapor elevates the boiling temperature and facilitates the cease
in boiling.\\
\hline 9&Influence of temperature&Threshold process, starts at a
certain temperature and terminates at another temperature, e.g.
balancing around a definite temperature takes place.&Threshold
process, starts at a certain temperature and terminates at another
temperature.\\
\hline 10&Working medium&Plasma&Mineral vacuum oil VM-1\\
\hline 11&Channel for energy drive out from the heated
surface&Energy if consumed on generation of strong magnetic
field.&Energy is consumed on vapor generation.\\
\hline
  \end{longtable}

One can see from this compendium that the phenomena of sunspot formation and
nucleate-free boiling in a thin layer have many common features: dynamics and conditions
of formation, geometry characteristics, structures on the surface, influence of the
formation site, periodicity. They also take the same place in the hierarchy of
self-organization for complex systems (rows 1-9 of the Table). The differences are in the
type of medium and the mechanism of energy channelling from the heating surface (rows 10,
11).

Pores and sunspots, as well as ``funnels'' with ``craters'' emerging during nucleate-free
boiling in a thin layer, develop in places with local overheating of surface. Beyond a
certain temperature (the saturation temperature) at the given pressure, the fluid
transforms into a metastable state. If the system has a local overheating above the
saturation temperature, this can be considered as an external impact. According to the Le
Chatelier principle, any equilibrium system being subjected to an external impact, tries
to change in a way that counteracts that impact. This means that in places of local
overheating the structures must be formed that provide removal of heat from the heating
surface and produce cooling of the heating surface. This creates pulsation in the surface
temperature. The process of funnel and crater formation in nucleate-free boiling in a
thin layer and creation of pores and sunspots are all the threshold-like process: it
starts after exceeding of a certain temperature and stops after cooling down to a certain
level, i.e., it exists in a narrow range of temperatures. The process is accompanied by
energy consumption and restructuring of flow pattern. Therefore, all the surface of a
spot, regardless its size, takes a uniform temperature (the model of protective shell
(Obridko \cite{Obr})).

The developed spot has the size multiple to the size of supergranulation, and all the
sunspots with intermediate sizes are much less stable (Obridko \cite{Obr}), because with
a nucleation of a temperature nonuniformity inside the Sun, it is ``carried away'' up to
the surface due to convective flow of solar within supergranules and giant granules.
While that, one or several neighbor supergranules must have a higher temperature. The
solar spot cools down the surface with the size multiple to the size of a supergranules.

As it has been pointed out, the difference between the formation of sunspots and surface
nucleate-free boiling in a thin layer of liquid is in the nature of working medium and
the method for removal of excessive energy from the heating surface (rows 10, 11 in the
Table.). Indeed, those two items are related tightly and reflect the fact that energy
transformation into another form is a complex process in these processes. For liquids,
the process of boiling means consumption of energy, generation of a new phase (vapor),
and an increase in volume. Simultaneously, the pressure of the vapor phase grows; for
vacuum boiling (explosive boiling) this fact is quite common, and a twofold pressure
growth can be observed.

During a nonstationary process of bubble formation (or crater) the local temperature of
the surface and liquid drops down and the volumetric pressure increases ($T_A\downarrow$,
$P_V\uparrow$). The pressure in volume has a direct correspondence with the boiling
temperature $T_S(P_V)$ that increases with the pressure, because there exists a relation
between the saturation temperature and the vapor pressure. Therefore, we have the
following: $T_A\downarrow , T_S(P_V)\uparrow$. Obviously, boiling (formation of bubbles
of craters) ceases at $T_A=T_S(P_V)$. Therefore, a decrease in the temperature of the
bubble (crater) formation and pressure growth {($~T_A\downarrow~,~P_V\uparrow$ )} must
inhibit boiling and transfer the liquid into equilibrium state for a time (during this
time, the bubble detaches or the crater collapses), and later the surface is heated up
again and the process repeats.

According to the very general principle formulated by Le
Chatelier, any system in equilibrium tends to rearrange for
weakening of the external impact. The heating of a patch of the
solar surface stimulates the process that must reduce the
temperature of this patch. The same way as liquid boiling channels
out the excessive energy on vapor generation, the solar energy is
consumed on generation of a magnetic field (this hypothesis was
put forward by Gurevich and Lebedinsky \cite{Gur,Leb}). The same
as in boiling liquids, the heated surface must be cooled down, and
this brings up the magnetic field pressure
\[
P_V(H)=\frac{H^2}{8\pi}
\]

That is, ($T_A\downarrow$  , $P_V (H)\uparrow$ ). If we consider
further this analogy, then the increasing magnetic field (similar
to the vapor pressure) suppresses the growth of a sunspot. This
analogy allows us to explain the accurate balancing of the
temperature and magnetic field strength around certain values.
However, this paper does not give the answer about an unambiguity
of the $P(H)-T$ relationship for solar substance (which is a
natural consequence of this analogy). But it is a known fact that
the spots can be observed on other stars, and they are similar to
sunspots. Many stars have the brightness of different levels and
the spot temperature and magnetic field strength must be different
from value of the Sun. This creates an opportunity to generalize
data in $P(H)-T$ coordinates (magnetic field strength -
brightness), but that kind of generalization is beyond the scope
of this article.

\section{CONCLUSION}

This paper put forward a new thermodynamic hypothesis of sunspot formation. An analogy
was considered between the processes of sunspot formation and phenomena of surface
nucleate-free boiling in a thin layer of liquid. This hypothesis goes in line with the Le
Chatelier principle and gives explanation why the brightness (temperature) of spots keeps
close to a certain level.

%\subsection*{Acknowledgment}
%I thanks D.A. Shapiro for their attention to Article and help in the
%LaTeX.

\newpage
\section*{List of captions}
\

Fig.1. Dynamics of ``crater'' formation: a - convective cell, b - ``funnel'', c -
``crater''.

Fig.2. Photos of the processes that were observed in the thin layer of oil VM -1 with the
thickness of 2 mm: a - ``funnel''; b - ``crater'' (shown with a cross); c - two
``craters'' with a bridge between them (arrow). Scale 1 cm.

Fig.3. Surface temperature pulsation during nucleate-free boiling in a thin layer  at low
heat flux.


\begin{thebibliography}{99}

\bibitem{Sit}Sitnik, G.F. 1939, Astronom. Zhurnal, v. 16, p.1.

\bibitem{Bray}Bray, R., Loughhead R. 1964, \emph {Sunspots} , Chapman and Hall Ltd.,
London.

\bibitem{Obr}Obridko, V.N. 1985, \emph {Sunspots and Complex of Activity}, Moscow, Nauka.

\bibitem{Gur}Gurevich, L.E., Lebedinsky, A.I. 1945, Sov. Phys. Doklady,  v. 49, p.
92.

\bibitem{Leb}Gurevich, L.E, Lebedinsky, A.I. 1946, JETP,  v. 16, p. 832, 840.

\bibitem{Moore}Moore, F.D., Mesler, R.B. 1961, AIChE Journal,  v.7, No. 4, p.620.

\bibitem{Kut}Kutateladze, S.S. 1979, \emph {Fundamentals of Heat Transfer}, Moscow,
Atomizdat.

\bibitem{Gog}Gogonin, I.I., Dorokhov, A.R., Zhukov, V.I. 1989, Izv. SO AN SSSR, Ser. Tehn. Nauk,  v.3, p.8. (in Russian, this Russian Journal Proceedings of Siberian Branch of the USSR Academy of Sciences/ Technical Sciences is translated completely into English as Soviet Journal of Applied Physics).

\bibitem{Zh96}Zhukov, V.I., 1996, Technical Physics Letters,  v.22, No. 21.

\bibitem{Zh99}Dorokhov, A.R., Zhukov, V.I. 1999, Journal of Engineering Physics and Thermophysics,
v. 72, No. 3, p. 430.

\bibitem{Ber}Berdnikov, V.S., Kirdyashkin, A.G. 1979, Bull. Sov. Acad.
Sci.  Ocean and Atmosph. Phys.,  v.15, No. 11,
p.1168.


\end{thebibliography}
\end{document}